\documentclass[manuscript]{aastex63}

\usepackage{color}
\usepackage{CJKutf8}
\newcommand{\cntext}[1]{\begin{CJK}{UTF8}{gbsn}#1\end{CJK}\kern-1ex}

\shortauthors{Kong et al.}
\begin{document}

\title{Dynamical modulation of solar flare electron acceleration due to plasmoid-shock interactions in the looptop region}

\correspondingauthor{Xiangliang Kong}
\email{kongx@sdu.edu.cn}

\author[0000-0003-1034-5857]{Xiangliang Kong (\cntext{孔祥良})}
\affiliation{Shandong Provincial Key Laboratory of Optical Astronomy and Solar-Terrestrial Environment,
and Institute of Space Sciences, Shandong University, Weihai, Shandong 264209, People's Republic of China}

\author[0000-0003-4315-3755]{Fan Guo (\cntext{郭帆})}
\affiliation{Los Alamos National Laboratory, Los Alamos, NM 87545, USA}
\affiliation{New Mexico Consortium, 4200 West Jemez Rd, Los Alamos, NM 87544, USA}

\author[0000-0002-9258-4490]{Chengcai Shen (\cntext{沈呈彩})}
\affiliation{Center for Astrophysics $\mid$ Harvard $\&$ Smithsonian, 60 Garden St, Cambridge, MA 02138, USA}

\author[0000-0002-0660-3350]{Bin Chen (\cntext{陈彬})}
\affiliation{Center for Solar-Terrestrial Research, New Jersey Institute of Technology,
323 Dr. Martin Luther King Blvd, Newark, NJ 07102, USA}

\author[0000-0001-6449-8838]{Yao Chen (\cntext{陈耀})}
\affil{Shandong Provincial Key Laboratory of Optical Astronomy and Solar-Terrestrial Environment,
and Institute of Space Sciences, Shandong University, Weihai, Shandong 264209, People's Republic of China}

\author{Joe Giacalone}
\affiliation{Department of Planetary Sciences, University of Arizona, Tucson, AZ 85721, USA}

\begin{abstract}

A fast-mode shock can form in the front of reconnection outflows and has been suggested as a promising site for particle acceleration in solar flares. Recent development of magnetic reconnection has shown that numerous plasmoids can be produced in a large-scale current layer.
Here we investigate the dynamical modulation of electron acceleration in the looptop region when plasmoids intermittently arrive at the shock by combining magnetohydrodynamics simulations with a particle kinetic model. As plasmoids interact with the shock, the looptop region exhibits various compressible structures that modulate the production of energetic electrons.
The energetic electron population varies rapidly in both time and space. The number of 5$-$10 keV electrons correlates well with the area with compression, while that of $>$50 keV electrons shows good correlation with strong compression area but only moderate correlation with shock parameters.
We further examine the impacts of the first plasmoid, which marks the transition from a quasi-steady shock front to a distorted and dynamical shock.
The number of energetic electrons is reduced by $\sim 20\%$ at 15$-$25 keV and nearly 40\% for 25$-$50 keV, while the number of 5$-$10 keV electrons increases.
In addition, the electron energy spectrum above 10 keV evolves softer with time.
We also find double or even multiple distinct sources can develop in the looptop region when the plasmoids move across the shock.
Our simulations have strong implications to the interpretation of nonthermal looptop sources, as well as the commonly observed fast temporal variations in flare emissions, including the quasi-periodic pulsations.

\end{abstract}



\section{Introduction} \label{sec:intro}

In solar flares, magnetic reconnection is believed to play a crucial role in the explosive release of magnetic energy in the corona \citep{shibata11,benz17}.
Observations have shown that an enormous number of particles are accelerated to high energies (up to tens of MeV for electrons) within tens of seconds to minutes and contain a significant fraction (as high as 10\%$-$50\%) of the released flare energy \citep{lin76,emslie12,aschwanden17}.
Various mechanisms may contribute to particle energization in flares, including
acceleration by parallel electric field, contracting/merging magnetic islands, or large-scale compression in the reconnection layer \citep[e.g.,][]{drake06,oka10,zhou15,li18a,li18b},
stochastic acceleration by magnetic turbulence or plasma waves \citep[e.g.,][]{miller96,petrosian04,Fu2020},
acceleration by the shrinkage of reconnected magnetic field lines \citep[e.g.,][]{somov97,karlicky06},
and acceleration by a fast-mode shock, often referred to as the flare termination shock (TS), driven by the reconnection outflows \citep[e.g.,][]{tsuneta98,mann09,warmuth09,guo12,li13,kong13,nishizuka13,chen15,kong19}.
However, it remains controversial which process plays a dominant role and can explain various observational signatures of particle energization in solar flares \citep{miller97,aschwanden02,zharkova11}.

Hard X-ray (HXR) and radio observations provide primary diagnostics of the acceleration and transport of energetic electrons in solar flares. Nonthermal looptop sources suggest that particle acceleration takes place above the top of flare loops and the TS is one of the promising candidates as the acceleration mechanism \citep[e.g.,][]{masuda94,melnikov02,krucker10,liu08,liu13,krucker14,oka15,gary18}. Although the TS has long been predicted in magnetohydrodynamic (MHD) simulations when fast reconnection outflows impinge upon the top of newly reconnected magnetic loops \citep[e.g.,][]{forbes86,magara96,yokoyama98}, there is rarely solid observational evidence. One of the reasons is that the spatial size of the TS is expected to be very limited \citep{chen19}.
In several studies, slow-drift radio emissions similar to Type II radio bursts (associated with shocks driven by coronal mass ejections) were interpreted as the radio signature of the TSs \citep{aurass02,mann09,warmuth09}.
Recently, using the Karl G. Jansky Very Large Array, \citet{chen15} presented a high-cadence radio spectroscopic imaging observation of stochastic spike bursts at decimetric wavelengths during a solar flare, which had a morphology and dynamic evolution most likely representative of a flare TS as suggested by numerical simulations. The observation suggested that the TS is located at the ending points of plasma downflows and slightly above a coronal HXR source. \citet{chen19} further performed detailed analysis of the split-band feature in the same radio spike burst event, and found that the high-frequency band is located slightly below the low-frequency band which, in turn, supported the interpretation that the two bands were emitted in the downstream and upstream side of the TS, respectively.

In addition to efficient particle acceleration, in some flare events, confinement of electrons in the looptop region is also required to account for the nonthermal looptop emissions \citep[e.g.,][]{simos13}.
Several mechanisms have been suggested, which include magnetic mirroring and turbulent pitch-angle scattering \citep[e.g.,][]{simos13,kontar14,musset18,ruan20}.
In theoretical models of particle acceleration in solar flares, the TS is often considered as a planar standing shock  \citep[e.g.,][]{tsuneta98,mann09,nishizuka13}.
However, recent MHD simulations have shown that the TS evolves dynamically and exhibits complex structures due to the impacts of jets/plasmoids \citep[e.g.,][]{takasao15,takasao16,takahashi17,shen18,cai19,zhao20}.
Particularly, a concave-downward magnetic structure has been shown to be present below the TS in the looptop region \citep{takasao15,shen18,kong19}.
Such a magnetic configuration is favorable for trapping electrons because it is more difficult for particles to travel transverse to the magnetic field than along it \citep{guo10,kong15,kong16}.

In order to investigate the role of the flare TSs in electron acceleration in more detail, however, there has been a lack of studies that couple a kinetic energetic-particle model with a realistic MHD simulation of solar flare region.
Recently, \citet{kong19} presented a model by numerically solving the Parker transport equation \citep{parker65} with the plasma velocity and magnetic field obtained from MHD simulations.
They found that the electron spectrum in the low energy resembles a power-law, which agreed well with the prediction of diffusive shock acceleration (DSA) theory.
The accelerated electrons are concentrated in the looptop region due to the acceleration at the TS and confinement by the magnetic trap structure, in agreement of HXR and microwave observations. Therefore, the model in \citet{kong19} can have strong implication to the acceleration and confinement of electrons in the nonthermal looptop sources.

During solar flares, fast temporal variations of flare emissions, including the so-called quasi-periodic pulsations (QPPs), are commonly observed in multiple wavelengths including radio and HXRs \citep{nakariakov09,McLaughlin18}. In addition, radio and HXR imaging occasionally show multiple sources at or above the flare looptops \citep[e.g.,][]{tomczak01,petrosian02,sui03,liu08,liu13,gary18,chen20,yu20}, suggesting more complicated dynamics during electron acceleration and transport.
\citet{kong19} explored electron acceleration and distribution in a quasi-steady and nearly symmetric phase of the TS. Recent theoretical work and numerical simulations have shown that a large-scale current sheet can break into numerous plasmoids \citep[e.g.,][]{shibata01,loureiro07,bhattacharjee09}.
As shown in \citet{shen18}, plasmoids are produced intermittently in the reconnection current sheet and interact dynamically with the TS. As a result, key properties of the TS, such as the compression ratio, Mach number and shock oblique angle, can significantly vary with time.

In this study, we investigate the dynamical modulation of electron acceleration and transport in the looptop region due to plasmoid-shock interactions.
The paper is structured as follows. We briefly introduce our numerical methods in Section \ref{sec:model} and present detailed analysis of the simulation results in Section \ref{sec:result}. We summarize and discuss the implications of this work in Section \ref{sec:conclusion}.

\section{Numerical Methods} \label{sec:model}
The numerical methods used in this study are nearly identical to those in \citet{kong19}. Here we provide salient details for completeness.
We first simulate the magnetic reconnection-driven TS in a classic two-ribbon solar flare geometry with a two-and-half dimensional (2.5-D) resistive MHD method \citep{chen15, shen18}, then we model the acceleration and transport of electrons in the looptop region using the MHD fields in a post-processing manner.

We perform a MHD simulation of a solar flare in the $x$$-$$y$ plane. The resistive MHD equations are solved with the Athena code \citep{stone08}.
The initial configuration is a vertical Harris-type current sheet.
The magnetic field lines are fixed to the bottom boundary by using a line-tied condition to model the two-ribbon flare. We use a uniform resistivity that gives the Lundquist number $S = 10^5$.
The normalization units are given as: the length $L_0$ = 75 Mm, the magnetic field $B_0$ = 40 G, the velocity $V_0$ = 810 km s$^{-1}$, and the time $t_0$ = 92 s.
Interested readers are referred to \citet{shen18} for details of the MHD setup and simulation results.

While \citet{kong19} has investigated electron acceleration and transport during a relatively steady period between 96.5$-$97.5 $t_0$, here we will pay special attention on a more dynamical evolution period during 97.4$-$98.5 $t_0$. To see the influence of plasmoid-shock interactions more clearly, we first model the acceleration of electrons for 3 $t_0$ (between 94.4$-$97.4 $t_0$ in Figure 2(a)) by using the fixed MHD frame at 97.4 $t_0$, which provides a quasi-steady distribution of energetic electrons before the plasmoids arrive. After 97.4 $t_0$, the particle simulation is performed with regard to the time-dependent MHD frames when the plasmoids interact with the flare TS.
We select the flare looptop region given by $x$ = [$-$0.15, 0.15] and $y$ = [0.4, 0.7] for particle simulation. The temporal cadence between adjacent MHD frames is 0.002 $t_0$, compared to 0.01 $t_0$ in \citet{kong19}, which better captures the dynamical evolution of the TS at shorter time scales.

Based on the plasma velocity and magnetic field from the MHD simulation, we model the acceleration and transport of electrons by numerically solving the Parker transport equation \citep{parker65}.
It is achieved by integrating stochastic differential equations corresponding to the Fokker$-$Planck form of the transport equation using a large number of pseudo-particles \citep[e.g.,][]{zhang99,giacalone08,guo10,kong17,li18b}.
The transport of charged particles in the magnetic field is described by the spatial diffusion coefficient. The diffusion coefficient parallel to the magnetic field is calculated from the qusai-linear theory \citep{jokipii71,giacalone99}, with an energy dependence $\kappa_{\parallel} = \kappa_{\parallel 0} (E/E_0)^{2/3}$. Following \citet{kong19}, we take $\kappa_{\parallel 0}$ = 0.005 $\kappa_0$, where $\kappa _0$ = $L_0 V_0$. We also consider perpendicular diffusion similar to results of test-particle simulations in synthetic turbulence \citep{giacalone99} and set $\kappa_\perp/\kappa_\parallel$ = 0.1.
In MHD simulations the TS can be resolved in several cells, which means that the shock width is on the order of one grid cell,  $\Delta x \sim$0.001 $L_0$. Therefore, in nearly all regions the characteristic diffusion length at the lowest energies is larger than the grid cell, $\kappa_{nn} /V_{0} > \Delta x$, where $\kappa_{nn}$ is the diffusion coefficient in the shock normal direction. We set the time step $\Delta t = 10^{-5}\ t_0$ at the injection energy to ensure that pseudo-particles can ``see'' the shock transition. 
\citet{kong19} has shown that with these parameters the electron spectrum in the low-energy range resembles a power-law, close to that predicted by the DSA theory.
Because the shock shape is very complex and evolves dynamically, here we inject particles uniformly in the particle simulation domain. The initial energy of electrons is fixed to be $E_0$ = 0.5 keV (corresponding to the typical energy of $\sim$6 MK plasma in the flaring region) and a total of 10$^8$ pseudo-particles are injected at a constant rate.

\section{Simulation Results} \label{sec:result}
In the MHD simulation, plasmoids form intermittently at the primary X-point at $y \sim$1.5 $L_0$ and move upward/downward surfing in the reconnection outflows. We examine three downward-moving plasmoids as they crash into the looptop region during 97.4--98.5 $t_0$ and interact with the TS.
Figures 1(a)--(b) show the maps of magnetic field strength $B$ and plasma velocity divergence $\nabla \cdot \textbf{V}$ at 97.4 $t_0$. The magnetic field is stronger in the plasmoids than that of the ambient, likely due to compression.
The TS locates at $y \sim$0.6 $L_0$, as well manifested by negative $\nabla \cdot \textbf{V}$ regions.
\citet{shen18} showed that the TS is a sharp transition layer, across which the flow speed decreases and the Mach number quickly drops from $\sim$2 to less than 1.
Figure 1(c) shows the time-distance plot of $\nabla \cdot \textbf{V}$ across the TS.
We find that the height of the TS varies dynamically. 
The average shock compression ratio $X$ ranges from 1.7 to 2.5, and shock angle $\theta_{Bn}$ ranges from 30$^{\circ}$ to 75$^{\circ}$ (see Figures 2(d)-(e)). As shown below, the TS can be deformed and restored, and multiple shocks/compression regions can be generated during the impacts of plasmoids.
\citet{kong19} has shown that electrons can be accelerated efficiently by the TS, leading to a concentrated electron population in the looptop region.
We expect that the dynamical plasmoid-shock interactions will affect both the shock properties and the acceleration of electrons.

Figure 2(a) shows temporal variations of the numbers of electrons at different energies, 5$-$10 keV, 15$-$25 keV, 25$-$50 keV, and $>$50 keV, respectively. The electron numbers are integrated over the looptop region, $x$ = [$-$0.1, 0.1] and $y$ = [0.5, 0.7], and normalized to their respective values at 97.4 $t_0$.
As noted in Section \ref{sec:model}, the MHD background before 97.4 $t_0$ is fixed. The number of energetic electrons at different energies rise gradually. Collectively, they develop a power-law energy spectrum, and reach a quasi-steady state prior to 97.4 $t_0$.
After 97.4 $t_0$, when the energetic particle simulations are performed with regard to the time-dependent MHD evolution, the three plasmoids collide with the TS successively with similar time intervals. The electron number profiles differ strongly at different energies.
Overall, after the series of plasmoid collisions, compared to the pre-collision steady-state values, the total number of low-energy (5--10 keV and 15--25 keV) electrons increases, but that of high-energy ($>$25 keV) electrons decreases.

The production of energetic electrons due to DSA depends on a number of factors such as the compression ratio, the shock angle, and magnetic field configuration in the loop-top region. Here we attempt to explore how the variation of number of energetic electrons depends on the properties of the TS. From the Parker equation we can find that the rate of particle acceleration is related to the magnitude of plasma compression, $\langle$$dp/dt$$\rangle$ $\sim$ $\langle$$- p \nabla \cdot \textbf{V}$$\rangle$, where $p$ is the particle momentum and $\textbf{V}$ is the flow velocity \citep{jokipii12}.
Figures 2(b)-(c) plot the temporal variation of the total number of cells with compression in the looptop region.
Compression area is illustrated by the number of cells where the velocity divergence $\nabla \cdot \textbf{V} <-$50 [$t_0^{-1}$], and we use a criterion of $\nabla \cdot \textbf{V} < -$150 [$t_0^{-1}$] for strong compression (mainly located at the TS).
The number of low-energy (5$-$10 keV) electrons correlates well with the total number of compression cells, with a correlation coefficient $cc$ = 0.88, and high-energy ($>$50 keV) electrons show good correlation with the total number of strong compression cells, with $cc$ = 0.70.
Figures 2(d)-(e) display the temporal variations of the shock compression ratio and shock angle averaged over the looptop region. There may be weak correlations with the $>$50 keV electrons ($cc$ = 0.38 and 0.49, respectively).
Note that here we only consider the primary shock at the front of downward reconnection outflows (i.e., the TS). As shown in Figures 4 and 5, multiple shocks/compression regions can form in the looptop region and contribute to particle energization.
We can only find moderately good correlations for high-energy electrons because they demand cruder conditions on various factors, such as the shock compression ratio and shock angle, and the background magnetic field configuration, and the acceleration time scale is longer than the variation time scales of these factors.

We find that the number of energetic electrons exhibit rapid variations at short ($<$0.1 $t_0$ or $<$9 s) time scales.
However, the variations during each plasmoid-shock interaction are not exactly the same.
This is because various factors, such as the size of plasmoids, the TS properties, and the magnetic field configuration, can affect the acceleration and transport of electrons.
In particular, the first plasmoid, denoted ``P1'' , is relatively small in size, which merges into the looptop region immediately after crossing the TS front. The size of the plasmoid is about $0.02\ L_0 = 1.5$ Mm, or about 2$''$ on the Sun.
\citet{chen15} showed that a temporary disruption of the TS by a fast plasma downflow with about the same size coincides with the reduction of the HXR and radio flux.
In this study, we mainly focus on examining the first plasmoid-shock interaction, which is comparable to the situation in \citet{chen15}.

The first plasmoid-shock interaction occurs between 97.5$-$97.85 $t_0$.
As shown in Figures \ref{fig:mhd_para_time}(b)--(e), at low energy of 5$-$10 keV, the electron number first decreases slightly and then increases gradually. At higher energies, the numbers decrease continually. The amount of decrease is more significant at higher energies. Eventually, the number of 15$-$25 keV electrons is reduced by $\sim 20\%$ and nearly 40\% for 25$-$50 keV. Both the trends and magnitudes are generally consistent with the evolution of X-ray flux observed by RHESSI and Fermi/GBM in \citet{chen15}.
Figure \ref{fig:spectrum}(a) shows the energy spectra of accelerated electrons at three times as marked by black arrows in Figure 2(a). The energy spectra below 10 keV are close to a power-law with a spectral index of $\delta$ = 2.5. The spectral index is consistent with the DSA prediction: if we take $X \approx 2$, a typical value for the flare TS as predicted in the MHD simulation \citep{shen18} and inferred from the observed split-band feature \citep{chen19}, the DSA predicts a power-law distribution with the spectral index $\delta = (X+2)/[2(X-1)] \approx 2$ in the non-relativistic limit. We note, however, in RHESSI observations, the X-ray emission from the low-energy part of the nonthermal electron spectrum (at $\lesssim$10--20 keV) is usually ``buried'' under that from flare-heated thermal plasma \citep{holman11}.  
The energy spectra above 20 keV deviate from a simple power-law and get softer with time.
Figure \ref{fig:spectrum}(b) shows a more detailed view of the spectra at $>$10 keV. At the energy range of several tens of keV, the degree of softening is $\sim$1. It is similar to that reported in \citet{chen15}, but we note that in their event the nonthermal ``tail'' is only observed (i.e., above the background) up to $\sim$25 keV, and they fitted the RHESSI X-ray spectra in the 10--25 keV range with a single power-law.

In Figure \ref{fig:plasmoid1}, the first column plots the maps of $\nabla \cdot \textbf{V}$, and the other two columns plot the spatial distributions of accelerated electrons at 5$-$10 keV and 25$-$50 keV, respectively.
Right before the collision, in panels (a)-(c), both low- and high-energy electrons are concentrated downstream of the TS, and the size at low energy is relatively larger, consistent with the results in our previous study \citep{kong19}.
Around 97.64 $t_0$, in panels (d)-(f), the centroid of plasmoid P1 is moving across the TS. The low-energy electrons are more uniformly distributed, and the number of 25$-$50 keV electrons has been largely reduced.
After the collision, in panels (g)-(i), the TS is distorted and separated into three parts, including one horizontal and two oblique shocks. Meanwhile, multiple shocks/compression regions form in the looptop region, possibly due to plasma backflow. Similar shocks have been discussed in previous MHD simulations \citep[e.g.,][]{takasao16,takahashi17,zhao20} and detailed analysis is beyond the scope of this work.
As discussed above, the number of 5$-$10 keV electrons increases, in accordance with the increase of compression cells. But the number of 25$-$50 keV electrons continually decreases.
\citet{chen15} showed that nonthermal electron population is reduced but not eliminated during the disruption of the flare TS that coincided with the arrival of a plasma downflow.
Our simulation results are consistent with their observations.

\begin{figure}
\centering
\includegraphics[width=0.95\linewidth]{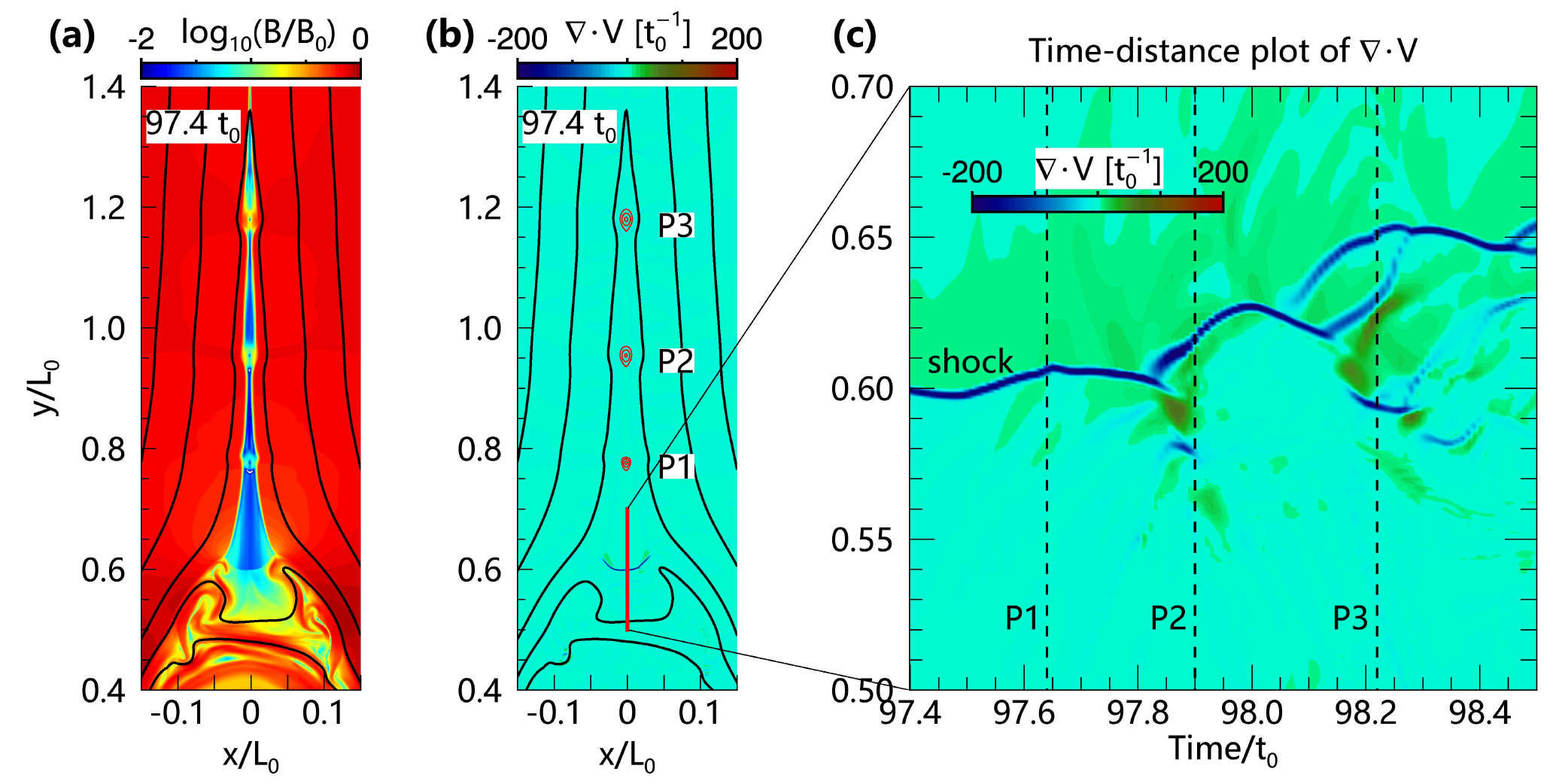}
\caption{
(a) and (b): Distributions of magnetic field strength $B$ and plasma velocity divergence $\nabla \cdot \textbf{V}$ at 97.4 $t_0$. The black curves show the magnetic field lines and the red contours of magnetic field in panel (b) denote the three plasmoids.
(c): Time-distance plot of $\nabla \cdot \textbf{V}$ across the TS between 97.4$-$98.5 $t_0$. Three vertical dashed lines mark the times when the plasmoid centroids encounter the TS.
}
\label{fig:mhd974}
\end{figure}

\begin{figure}
\centering
\includegraphics[width=0.95\linewidth]{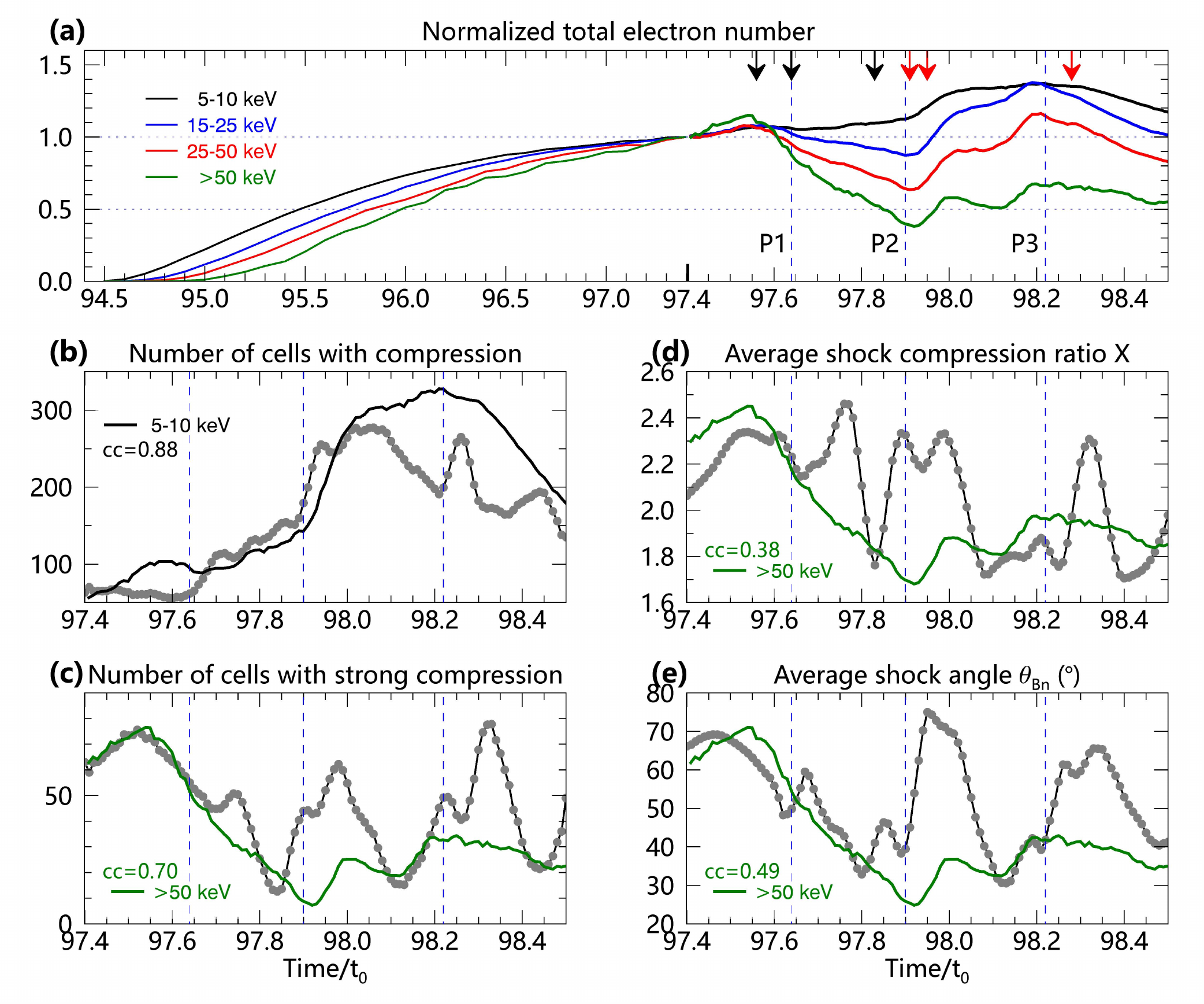}
\caption{
(a) Temporal variations of the total number of electrons at a variety of energy ranges integrated over the looptop region and normalized to their respective values at 97.4 $t_0$.
(b)-(c): Temporal variations of the total number of (strong) compression cells in the looptop region, with the criterions of the velocity divergence $\nabla \cdot \textbf{V}$  $<-$50 and $<-$150 [$t_0^{-1}$], respectively.
(d)-(e): Temporal variations of the shock density compression ratio $X$ and shock angle $\theta_{Bn}$ averaged over the looptop region.
Three blue vertical dashed lines in each panel mark the times when the three plasmoids cross the TS.
}
\label{fig:mhd_para_time}
\end{figure}

\begin{figure}
\centering
\includegraphics[width=0.95\linewidth]{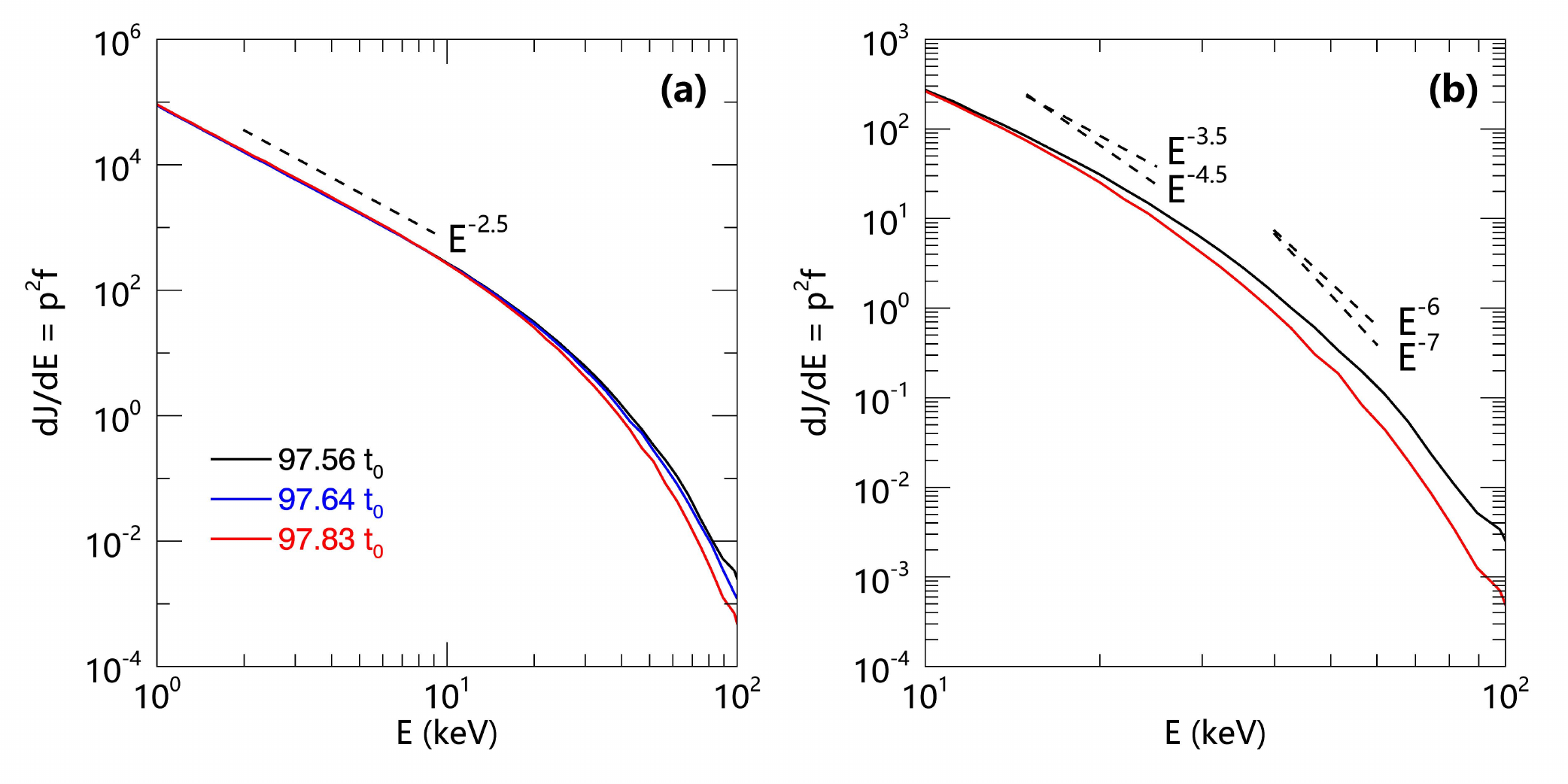}
\caption{
(a) Energy spectra of accelerated electrons during the first plasmoid-shock interaction at three times (before the plasmoid encountering the TS at 97.56 $t_0$, when its centroid crossing the TS at 97.64 $t_0$, and before the second plasmoid encountering the TS at 97.83 $t_0$), as marked by black arrows in Figure 2(a). (b) Energy spectra at 97.56 and 97.83 $t_0$ zoom-in to the high-energy range.
}
\label{fig:spectrum}
\end{figure}

\begin{figure}
\centering
\includegraphics[width=0.95\linewidth]{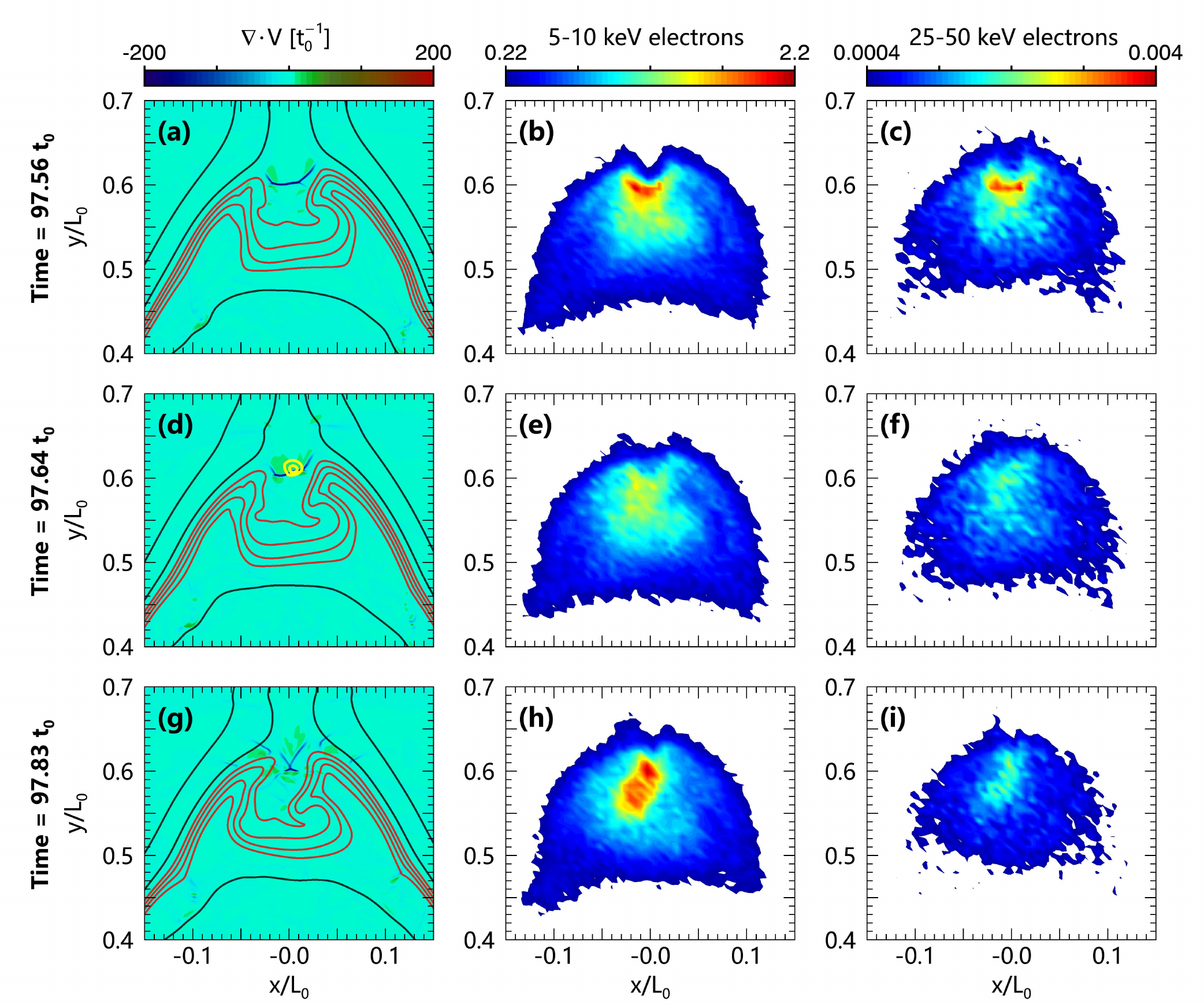}
\caption{
The first column shows the maps of velocity divergence $\nabla \cdot \textbf{V}$ from MHD simulation. The red field lines plot the magnetic trap in the looptop region, and the yellow contours denote the plasmoid P1.
The other two columns show the spatial distributions of accelerated electrons at 5$-$10 keV and 25$-$50 keV.
The time in each row is the same and marked by black arrows in Figure 2(a).
}
\label{fig:plasmoid1}
\end{figure}

\begin{figure}
\centering
\includegraphics[width=0.95\linewidth]{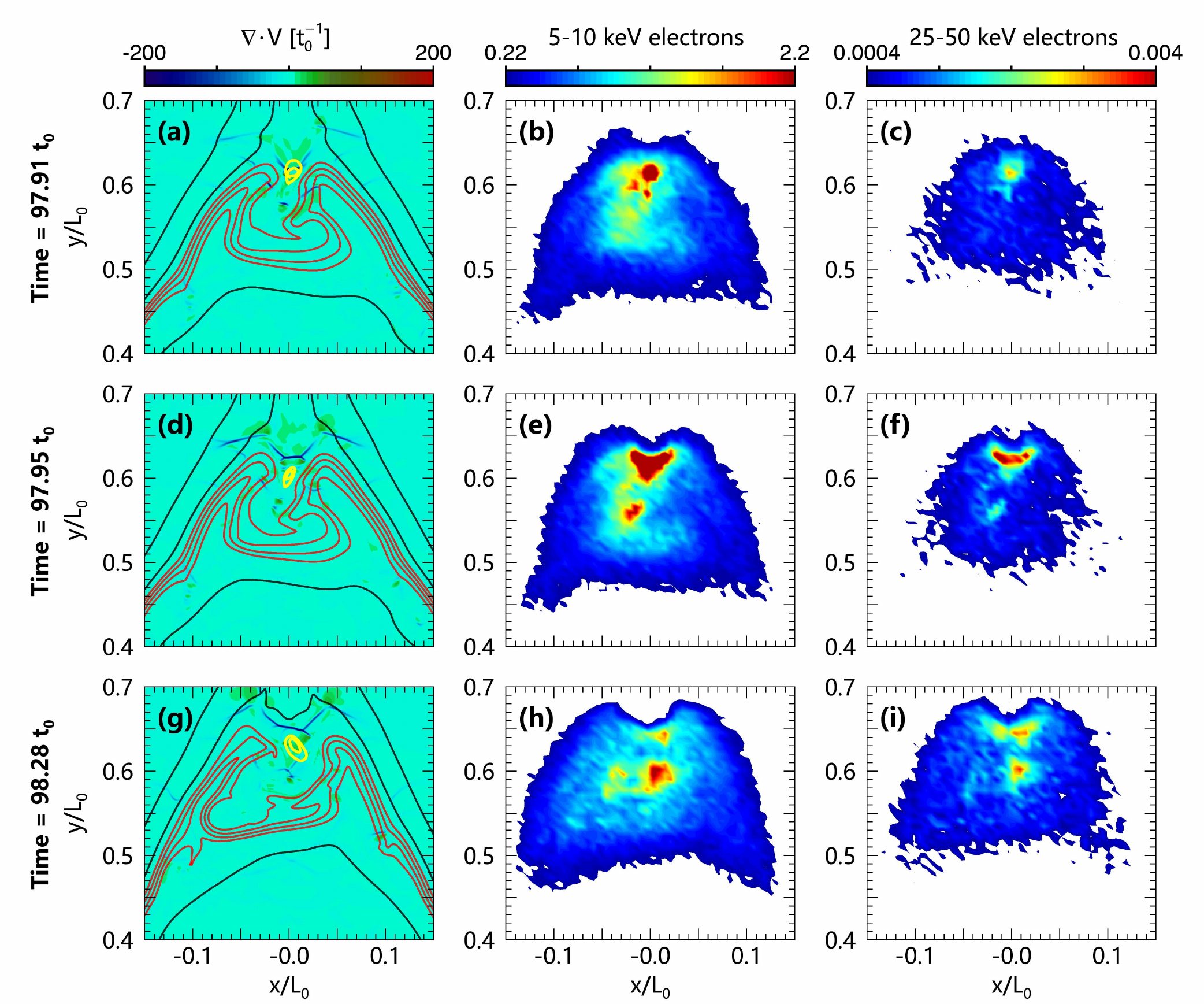}
\caption{
Same plots as in Figure 4. The three times are marked by red arrows in Figure 2(a). The yellow contours denote the plasmoid P2 in panels (a) and (d), and the plasmoid P3 in panel (g).
}
\label{fig:plasmoid23}
\end{figure}

We now analyze the effects of the other two plasmoids, P2 and P3.
In Figure \ref{fig:plasmoid23}, the first two rows show the distributions of accelerated electrons during the impact of plasmoid P2. In the course of P2 moving across the TS, a very compact source appears both at low- and high- energies.
When P2 reaches the downstream region, an intense source is distributed around the TS, similar to the case at 97.56 $t_0$ in Figure \ref{fig:plasmoid1}. As shown in Figure \ref{fig:mhd_para_time}(d) and (e), during 97.9$-$98.0 $t_0$, the TS has a nearly perpendicular shock geometry and a high compression ratio, both of which are favorable for efficient particle acceleration. This explains why the electron numbers increase when P2 moves across the TS.
We can also find another smaller source at $y \sim$0.55 $L_0$. The distance between the two sources is $\sim$0.06 $L_0$ $\sim$5 Mm.
Later, during 98.0$-$98.1 $t_0$, the horizontal shock is deformed. As shown in Figure \ref{fig:mhd_para_time}, both the shock compression ratio and shock angle decrease. As a result, the growth rate of electron number becomes slower, and the number of $>$50 keV electrons even is reduced.
In the bottom row, it shows that multiple sources appear after the collision of plasmoid P3.
The upper source is located near the TS. The lower source is located at $\sim$0.6 $L_0$, which coincides with a compression region existing for nearly 0.1 $t_0$, as shown in Figure \ref{fig:mhd974}(c).

\section{Conclusions and Discussion} \label{sec:conclusion}
In this study, we perform a numerical simulation of electron acceleration and transport in the flare looptop region. We explore the impacts of plasmoids that intermittently crash into the TS. As a result, the TS evolves dynamically and can be distorted and restored.
We find that the energetic electron population shows a rapid variation in both time and space, with distinct differences at low and high energies.
The number of low-energy (5--10 keV) electrons correlates strongly with the total area of compression regions, while the high-energy ($>$50 keV) electrons show good correlation with strong compression cells but moderate correlations with the shock compression ratio and shock angle.

We have focused on studying the impacts of the first plasmoid-shock interaction, when the shock front first experiences a transition from a quasi-steady state to a highly dynamic period. The simulation results during this interaction compare favorably with the observations reported in \citet{chen15} regarding the evolution of the TS front (outlined by the radio spike centroids) as well as the associated X-ray and radio emissions.
The plasmoid merges into the looptop region immediately after crossing the TS.
The number of 15$-$25 keV electrons is reduced by $\sim 20\%$ and nearly 40\% for 25$-$50 keV, while the number of 5$-$10 keV electrons increases.
The spatial distributions of accelerated electrons also show that the number of 25$-$50 keV electrons has been greatly reduced throughout the looptop region.
In addition, in accordance with the observations of the X-ray spectra reported in \citet{chen15}, the energy spectra above 10 keV evolves softer in time immediately following the arrival of plasmoid P1.
During the subsequent two plasmoid-shock interactions, both the electron number profiles and spatial distributions display a more complicated temporal and spatial evolution, with multiple sources appear when the plasmoids interact and move across the TS front.

In solar flares, QPPs are ubiquitously observed in almost all wavelengths including HXR and radio emissions \citep{nakariakov09,McLaughlin18}.
The typical periods range from a fraction of a second to serval minutes \citep[e.g.,][]{tan10,yuan19,li20}, therefore possibly related to processes in the MHD regime.
%Quasi-periodic oscillations have also been found in the looptop region in MHD simulations \citep[e.g.,][]{takasao16,takahashi17,cai19}.
In our simulations, in response to the arrival of the three plasmoids, the total electron number at different energies shows a quasi-periodic temporal variation with a time interval of tens of seconds.
Although the limited number of plasmoid interactions in our simulation renders it difficult to directly associate the temporal evolution of the energetic electrons (and the nonthermal emissions) to the so-called QPPs, our results suggest that the spatial and energy distribution of the electrons in the looptop region can indeed exhibit rapid temporal evolution in response to the arrival of the plasmoids at the TS front. If the formation of the plasmoids exhibits a quasi-periodic behavior, as was demonstrated in a number of numerical studies \citep[e.g.,][]{shen11}, our results may provide a viable explanation for the formation of the QPPs. We note that this scenario was previously suggested by \citet{takasao16}, although they did not have particle simulations to explicitly show the temporal and spatial evolution of energetic electrons in the vicinity of the flare TS.
%The formation of plasmoids mainly depends on the factors  such as the plasma $\beta$ and the Lundquist number.
%Therefore, it is of particular importance to investigate the effects of those factors on the evolution of the TS and electron acceleration in the future.
The periodicity in our simulation is highly correlated with the production rate of plasmoids in the flare current sheet. Both in linear theory \citep{loureiro07} and 2D MHD simulations \citep[e.g.,][]{cassak09,samtaney09,huang10}, it has been shown that the number of plasmoids formed in the reconnection current sheet depends on the value of Lundquist number. If the Lundquist number is much larger, as in the realistic solar corona, many more but smaller plasmoids can be generated. The size (magnetic flux) distribution of plasmoids has also been investigated \citep[e.g.,][]{fermo10,uzdensky10,huang12,shen13,ni15}. Therefore, the value of Lundquist number could have important effect on the properties of the TS, the acceleration of energetic electrons and the periodicity in flare emissions. This will be explored in detail in our future work. In addition, the structure in the current layer has shown to be even more complex in 3D numerical simulations \citep[e.g.,][]{guo15,yang20}. However, it remains unknown how the 3D turbulence and physics influence the physical properties and dynamical evolution of the flare TS.

X-ray and microwave emissions in the corona occasionally show multiple sources \citep[e.g.,][]{tomczak01,petrosian02,sui03,liu08,liu13,yu20}.
Particularly, the double coronal X-ray sources often display energy-dependent feature, i.e., the centroid locations becoming closer at higher energies.
The distance between the upper and lower sources range from a few arcseconds to tens of arcseconds and the sources can be visible for several minutes and even longer.
The double X-ray sources are attributed to particle acceleration taking place both in the upward and downward reconnection outflow regions \citep{liu13}.
MHD simulations have shown that TSs can be generated both at the flare looptop and the bottom of the flux rope \citep[e.g.,][]{takahashi17,zhao20}.
In our work, we have only focused on the lower TS in the looptop region. The distance between the upper and lower sources is a few Mm and only lasts for a few seconds. Therefore, it cannot explain the double coronal X-ray sources as shown in \citet{liu13}, but predicts that multiple sources can occur sporadically in the looptop region due to the moudulations in electron acceleration and transport when jets/plasmoids interact with the TS. We note that a recent study by \citet{yu20} does show such a double X-ray (and microwave) source in the looptop region during the post-impulsive phase of the 2017 September 10 X8.2 flare.

There has been lacking studies that couple a kinetic energetic-particle model with a realistic MHD simulation of solar flare region.
We present a model by numerically solving the Parker transport equation with the plasma velocity and magnetic field from MHD simulations.
Our simulations can reproduce the energy spectrum and spatial distributions of energetic electrons necessary for explaining the nonthermal looptop sources.
In our future work, we will combine our MHD-particle model with radiation models to investigate the dynamical evolution in HXR and microwave emissions during solar flares.

\acknowledgments
The authors thank Xiaocan Li and Sijie Yu for helpful discussions.
This work was supported by the National Natural Science Foundation of China under grants 11873036, 42074203 and 11790303 (11790300), the Young Elite Scientists Sponsorship Program by China Association for Science and Technology, and the Young Scholars Program of Shandong University, Weihai.
F.G. is supported by NSF grant AST-1735414 and DOE grant DE-SC0018240.
B.C. acknowledges support by NSF grant AST-1735405 to NJIT.
The work was carried out at National Supercomputer Center in Guangzhou (TianHe-2).



%\appendix

%\section{Appendix information}


%\bibliography{sample63}{}

\begin{thebibliography}{}
\bibitem[Aschwanden(2002)]{aschwanden02}Aschwanden, M. J. 2002, SSRv, 101, 1
\bibitem[Aschwanden et al.(2017)]{aschwanden17}Aschwanden, M. J., Caspi, A. Cohen, C. M. S., et al. 2017, ApJ, 836, 17
\bibitem[Aurass et al.(2002)]{aurass02}Aurass, H., Vr$\check{s}$nak, B., \& Mann, G. 2002, A\&A, 384, 273
%\bibitem[Aurass \& Mann(2004)]{aurass04}Aurass, H., \& Mann, G. 2004, ApJ, 615, 526
\bibitem[Bhattacharjee et al.(2009)]{bhattacharjee09}Bhattacharjee, A., Huang, Y.-M., Yang, H., \& Rogers, B. 2009, PhPl, 16, 112102
\bibitem[Benz(2017)]{benz17}Benz, A. O. 2017, LRSP, 14, 2
\bibitem[Cai et al.(2019)]{cai19}Cai, Q., Shen, C., Raymond, J. C., et al. 2019, MNRAS, 489, 3183
\bibitem[Cassak et al.(2009)]{cassak09}Cassak, P. A., Shay, M. A., \& Drake, J. F. 2009, PhPl, 16, 120702
\bibitem[Chen et al.(2015)]{chen15}Chen, B., Bastian, T. S., Shen, C., et al. 2015, Sci, 350, 1238
\bibitem[Chen et al.(2019)]{chen19}Chen, B., Shen, C., Reeves, K.~K., Guo, F., \& Yu, S.\ 2019, ApJ, 884, 63
\bibitem[Chen et al.(2020)]{chen20}Chen, B., Shen, C., Gary, D.~E., et al.\ 2020, Nature Astronomy, doi:10.1038/s41550-020-1147-7
%\bibitem[Daughton et al.(2009)]{daughton09}Daughton, W., Roytershteyn, V., Albright, B. J., et al. 2009, PhRvL, 103, 065004
\bibitem[Drake et al.(2006)]{drake06}Drake, J. F., Swisdak, M., Che, H., \& Shay, M. A. 2006, Natur, 443, 553
\bibitem[Emslie et al.(2012)]{emslie12}Emslie, A. G., Dennis, B. R., Shih, A. Y., et al. 2012, ApJ, 759, 71
\bibitem[Fermo et al.(2010)]{fermo10}Fermo, R. L., Drake, J. F., \& Swisdak, M. 2010, PhPl, 17, 010702
\bibitem[Forbes(1986)]{forbes86}Forbes, T. G. 1986, ApJ, 305, 553
%\bibitem[Forbes(1988)]{forbes88}Forbes, T. G. 1988, SoPh, 117, 97
\bibitem[Fu et al.(2020)]{Fu2020} Fu, X., Guo, F., Li, H., et al.\ 2020, \apj, 890, 161
\bibitem[Gary et al.(2018)]{gary18} Gary, D.~E., Chen, B., Dennis, B.~R., et al.\ 2018, \apj, 863, 83
\bibitem[Giacalone \& Jokipii(1999)]{giacalone99}Giacalone, J., \& Jokipii, J. R.\ 1999, \apj, 520, 204
\bibitem[Giacalone \& Neugebauer(2008)]{giacalone08} Giacalone, J., \& Neugebauer, M.\ 2008, \apj, 673, 629
%\bibitem[Guo \& Giacalone(2010)]{guo10} Guo, F., \& Giacalone, J.\ 2010, \apj, 715, 406
\bibitem[Guo \& Giacalone(2012)]{guo12}Guo, F., \& Giacalone, J. 2012, ApJ, 753, 28
\bibitem[Guo et al.(2015)]{guo15}Guo, F., Liu, Y.-H., Daughton, W., \& Li, H. 2015, ApJ, 806, 167
\bibitem[Guo et al.(2010)]{guo10}Guo, F., Jokipii, J. R., \& Kota, J. 2010, ApJ, 725, 128
\bibitem[Holman et al.(2011)]{holman11}Holman, G. D., Aschwanden, M. J., Aurass, H., et al. 2011, SSRv, 159, 107
\bibitem[Huang \& Bhattacharjee(2010)]{huang10}Huang, Y.-M., \& Bhattacharjee, A. 2010, PhPl, 17, 062104
\bibitem[Huang \& Bhattacharjee(2012)]{huang12}Huang, Y.-M., \& Bhattacharjee, A. 2012, PhRvL, 109, 265002
\bibitem[Jokipii(1971)]{jokipii71}Jokipii, J. R. 1971, Rev. Geophys., 9, 27
\bibitem[Jokipii(2012)]{jokipii12}Jokipii, J. R. 2012, AIPC, 1436, 144
\bibitem[Karlick\'{y} \& B\'{a}rta(2006)]{karlicky06}Karlick\'{y}, M., \& B\'{a}rta, M. 2006, ApJ, 647, 1472
\bibitem[Kong et al.(2015)]{kong15}Kong, X., Chen, Y., Guo, F., et al.\ 2015, \apj, 798, 81
\bibitem[Kong et al.(2016)]{kong16}Kong, X., Chen, Y., Guo, F., et al.\ 2016, \apj, 821, 32
\bibitem[Kong et al.(2017)]{kong17}Kong, X., Guo, F., Giacalone, J., Li, H., \& Chen, Y. 2017, \apj, 851, 38
%\bibitem[Kong et al.(2019)]{kong19} Kong, X., Guo, F., Chen, Y., \& Giacalone, J.\ 2019, \apj, 883, 49
\bibitem[Kong et al.(2019)]{kong19} Kong, X., Guo, F., Shen, C.,  et al.\ 2019, \apjl, 887, L37
\bibitem[Kong et al.(2013)]{kong13}Kong, X., Li, G., \& Chen, Y. 2013, ApJ, 774, 140
\bibitem[Kontar et al.(2014)]{kontar14}Kontar, E. P., Bian, N. H., Emslie, A. G., \& Vilmer, N. 2014, ApJ, 780, 176
\bibitem[Krucker \& Battaglia(2014)]{krucker14}Krucker, S., \& Battaglia, M. 2014, ApJ, 780, 107
%\bibitem[Krucker et al.(2008)]{krucker08}Krucker, S., Battaglia, M., Cargill, P. J., et al. 2008, A\&ARv, 16, 155
\bibitem[Krucker et al.(2010)]{krucker10}Krucker, S., Hudson, H. S., Glesener, L., et al. 2010, ApJ, 714, 1108
\bibitem[Li et al.(2013)]{li13}Li, G., Kong, X., Zank, G., \& Chen, Y. 2013, ApJ, 769, 22
\bibitem[Li et al.(2018a)]{li18a}Li, X., Guo, F., Li, H., \& Birn, J.\ 2018a, \apj, 855, 80
\bibitem[Li et al.(2018b)]{li18b}Li, X., Guo, F., Li, H., \& Li, S. 2018b, ApJ, 866, 4
\bibitem[Li et al.(2020)]{li20}Li, D., Kolotkov, D., Nakariakov, V., Lu, L., \& Ning, Z. 2020, \apj, 888, 53
\bibitem[Lin \& Hudson(1976)]{lin76}Lin, R. P., \& Hudson, H. S. 1976, Sol. Phys., 50, 153
\bibitem[Liu et al.(2008)]{liu08}Liu, W., Petrosian, V., Dennis, B. R., \& Jiang, Y. W. 2008, ApJ, 676, 704
\bibitem[Liu et al.(2013)]{liu13}Liu, W., Chen, Q., \& Petrosian, V. 2013, ApJ, 767, 168
\bibitem[Loureiro et al.(2007)]{loureiro07}Loureiro, N. F., Schekochihin, A. A., \& Cowley, S. C. 2007, PhPl, 14, 100703
\bibitem[McLaughlin et al.(2018)]{McLaughlin18}McLaughlin, J. A., Nakariakov, V. M., Dominique, M., Jelínek, P., \& Takasao, S. 2018, SSRv, 214, 45
\bibitem[Magara et al.(1996)]{magara96}Magara, T., Mineshige, S., Yokoyama, T. and Shibata, K., 1996, ApJ, 466, 1054
\bibitem[Mann et al.(2009)]{mann09}Mann, G., Warmuth, A., \& Aurass, H. 2009, A\&A, 494, 669
\bibitem[Masuda et al.(1994)]{masuda94}Masuda, S., Kosugi, T., Hara, H., Tsuneta, S., \& Ogawara, Y. 1994, Natur, 371, 495
\bibitem[Melnikov et al.(2002)]{melnikov02} Melnikov, V.~F., Shibasaki, K., \& Reznikova, V.~E.\ 2002, \apjl, 580, L185
\bibitem[Miller et al.(1997)]{miller97}Miller, J. A., Cargill, P. J., Emslie, A. G., et al. 1997, JGR, 102, 14631
\bibitem[Miller et al.(1996)]{miller96}Miller, J. A., Larosa, T. N., \& Moore, R. L. 1996, ApJ, 461, 445
\bibitem[Musset et al.(2018)]{musset18}Musset, S., Kontar, E. P., \& Vilmer, N. 2018, A\&A, 610, A6
\bibitem[Nakariakov \& Melnikov(2009)]{nakariakov09}Nakariakov, V. M., \& Melnikov, V. F. 2009, SSRv, 149, 119
\bibitem[Ni et al.(2015)]{ni15}Ni, L., Lin, J., Mei, Z., \& Li, Y. 2015, ApJ, 812, 92
\bibitem[Nishizuka \& Shibata(2013)]{nishizuka13}Nishizuka, N. \& Shibata, K. 2013, PhRvL, 110, 051101
%\bibitem[Oka et al.(2018)]{oka18}Oka, M., Birn, J., Battaglia, M., et al. 2018, SSRv, 214, 82
\bibitem[Oka et al.(2015)]{oka15}Oka, M., Krucker, S., Hudson, H. S., Saint-Hilaire, P. 2015, ApJ, 799, 129
\bibitem[Oka et al.(2010)]{oka10}Oka, M., Phan, T. D., Krucker, S., et al. 2010, ApJ, 714, 915
\bibitem[Parker(1965)]{parker65}Parker, E.~N.\ 1965, \planss, 13, 9
\bibitem[Petrosian et al.(2002)]{petrosian02}Petrosian, V., Donaghy, T. Q., \& McTiernan, J. M. 2002, ApJ, 569, 459
\bibitem[Petrosian \& Liu(2004)]{petrosian04}Petrosian, V., \& Liu, S. 2004, ApJ, 610, 550
\bibitem[Ruan et al.(2020)]{ruan20}Ruan, W., Xia, C., \& Keppens, R.\ 2020, \apj, 896, 97
\bibitem[Samtaney et al.(2009)]{samtaney09}Samtaney, R., Loureiro, N. F., Uzdensky, D. A., Schekochihin, A. A., \& Cowley, S. C. 2009, PhRvL, 103, 105004
%\bibitem[Seaton \& Forbes(2009)]{seaton09}Seaton, D. B., \& Forbes, T. G. 2009, ApJ, 701, 348
\bibitem[Sim\~{o}es \& Kontar(2013)]{simos13}Sim\~{o}es, P. J. A., \& Kontar, E. P. 2013, A\&A, 551, A135
\bibitem[Shen et al.(2018)]{shen18}Shen, C., Kong, X., Guo, F., Raymond, J. C., \& Chen, B. 2018, ApJ, 869, 116
\bibitem[Shen et al.(2011)]{shen11}Shen, C., Lin, J., \& Murphy, N. A. 2011, ApJ, 737, 14
\bibitem[Shen et al.(2013)]{shen13}Shen, C., Lin, J., Murphy, N. A., \& Raymond, J. C. 2013, PhPl, 20, 072114
\bibitem[Shibata \& Magara(2011)]{shibata11}Shibata, K., \& Magara, T. 2011, LRSP, 8, 6
\bibitem[Shibata et al.(1995)]{shibata95}Shibata, K., Masuda, S., Shimojo, M., et al. 1995, ApJL, 451, L83
\bibitem[Shibata \& Tanuma(2001)]{shibata01}Shibata, K., \& Tanuma, S. 2001, EP\&S, 53, 473
\bibitem[Somov \& Kosugi(1997)]{somov97}Somov, B. V., \& Kosugi, T. 1997, ApJ, 485, 859
%\bibitem[Su et al.(2013)]{su13}Su, Y., Veronig, A. M., Holman, G. D., et al. 2013, NatPh, 9, 489
\bibitem[Sui \& Holman(2003)]{sui03}Sui, L., \& Holman, G. D. 2003, ApJL, 596, L251
\bibitem[Stone et al.(2008)]{stone08}Stone, J. M., Gardiner, T. A., Teuben, P., Hawley, J. F., \& Simon, J. B. 2008, ApJS, 178, 137
\bibitem[Tan et al.(2010)]{tan10}Tan, B., Zhang, Y., Tan, C., \& Liu, Y. 2010, ApJ, 723, 25
\bibitem[Takahashi et al.(2017)]{takahashi17}Takahashi, T., Qiu, J., \& Shibata, K. 2017, ApJ, 848, 102
\bibitem[Takasao et al.(2015)]{takasao15}Takasao, S., Matsumoto, T., Nakamura, N., \& Shibata, K. 2015, ApJ, 805, 135
\bibitem[Takasao \& Shibata(2016)]{takasao16}Takasao, S., \& Shibata, K. 2016, ApJ, 823, 150
\bibitem[Tomczak(2001)]{tomczak01}Tomczak, M. 2001, A\&A, 366, 294
\bibitem[Tsuneta \& Naito(1998)]{tsuneta98}Tsuneta, S., \& Naito, T. 1998, ApJL, 495, L67
\bibitem[Uzdensky et al.(2010)]{uzdensky10}Uzdensky, D. A., Loureiro, N. F., \& Schekochihin, A. A. 2010, PhRvL, 105, 235002
%\bibitem[Van Doorsselaere et al.(2016)]{vanDoorsselaere16}Van Doorsselaere, T., Kupriyanova, E. G., \& Yuan, D. 2016, SoPh, 291, 3143
\bibitem[Warmuth et al.(2009)]{warmuth09}Warmuth, A., Mann, G., \& Aurass, H. 2009, A\&A, 494, 677
%\bibitem[Wu et al.(2016)]{wu16}Wu, Z., Chen, Y., Huang, G., et al. 2016, ApJL, 820, L29
\bibitem[Yang et al.(2020)]{yang20}Yang, L., Li, H., Guo, F., et al. 2020, ApJL, 901, L22
\bibitem[Yokoyama \& Shibata(1998)]{yokoyama98}Yokoyama, T., \& Shibata, K. 1998, ApJL, 494, L113
%\bibitem[Yokoyama \& Shibata(2001)]{yokoyama01}Yokoyama, T., \& Shibata, K. 2001, ApJ, 549, 1160
\bibitem[Yu et al.(2020)]{yu20}Yu, S., Chen, B., Reeves, K. K., et al. 2020, \apj, 900, 17
\bibitem[Yuan et al.(2019)]{yuan19}Yuan, D., Feng, S., Li, D., Ning, Z., \& Tan, B. 2019, ApJL, 886, L25
\bibitem[Zhang(1999)]{zhang99}Zhang, M. 1999, ApJ, 513, 409
\bibitem[Zhao \& Keppens(2020)]{zhao20}Zhao, X., \& Keppens, R.\ 2020, \apj, 898, 90
%\bibitem[Zhao et al.(2019)]{zhao19}Zhao, X., Xia, C., Van Doorsselaere, T., Keppens, R., \& Gan, W. 2019, ApJ, 872, 190
\bibitem[Zharkova et al.(2011)]{zharkova11}Zharkova, V. V., Arzner, K., Benz, A. O., et al. 2011, SSRv, 159, 357
\bibitem[Zhou et al.(2015)]{zhou15}Zhou, X., B\"{u}chner, J., B\'{a}rta, M., Gan, W., \& Liu, S. 2015, ApJ, 815, 6
\end{thebibliography}
%\bibliographystyle{aasjournal}

\end{document}